# Behavior of Wireless Body-to-Body Networks Routing Strategies for Public Protection and Disaster Relief

Dhafer Ben Arbia[1,2], Muhammad Mahtab Alam[1], Rabah Attia[2], Elyes Ben Hamida[1]

[1] Qatar Mobility Innovations Center (QMIC), Qatar Science and Technology Park (QSTP), PO Box 210531, Doha, Qatar.
[2] SERCOM Lab, Polytechnic School of Tunisia, University of Carthage B.P. 743. 2078 La Marsa. Tunisia.
Email : dhafera@qmic.com, mahtaba@qmic.com, rabah.attia@enit.rnu.tn, elyesb@qmic.com

*Abstract*— Critical and public safety operations require real-time data transfer from the incident area(s) to the distant operations command center going through the evacuation and medical support areas. Any delay in communication may cause significant loss. In some cases, it is anticipated that the existing communication infrastructures can be damaged or out-of-service. It is thus required to deploy tactical ad-hoc networks to cover the operation zones. Routing data over the deployed network is a significant challenge with consideration to the operations conditions. In this paper we evaluate the performance of mutli-hop routing protocols while using different wireless technologies in an urban critical and emergency scenario. Using a realistic mobility model, Mobile Ad hoc, geographic based and data-centric routing protocols are evaluated with different communication technologies (*i.e.* WiFi IEEE 802.11; WSN IEEE 802.15.4; WBAN IEEE 802.15.6). It is concluded that, WiFi IEEE 802.11 is the best wireless technology with consideration to the packet reception rate and the energy consumption. Whereas, in terms of delay, WBAN IEEE 802.15.6 is the most efficient. With regards to the routing protocols, assuming that the location information is available, geographical based routing protocol with WiFi IEEE 802.11 performed much better compared to the others routing protocols. In case where the location information is unavailable, gradient based routing protocol with WBAN IEEE 802.15.6 seems the best combination.

## I. INTRODUCTION

Disasters are increasing worldwide, impacting not only the economies and infrastructure but significant number of human lives. With regards to the emergency response to these disasters, the role of effective *Public Safety Network* (PSN) infrastructures (*e.g.* TETRA, LTE) is extremely vital. However, it is anticipated that, during and after the disasters, existing PSN infrastructures can be either completely damaged or oversaturated. Consequently, there is a growing demand for ubiquitous emergency response system which could be easily and rapidly deployed.

In this context, it is expected that *Wearable Body Area Networks* (W-BAN) can play a key role in not only providing an additional connectivity network but it can also be used to monitor the physiological status of the involved workforces and the surrounding environment [1]. W-BAN consists of smart, low-power, miniaturized devices such as sensors (*i.e.*, to sense, transmit and receive data), actuator (to act on the perceive data) and coordinator (to act as a gateway to external networks). W-BAN communication architecture is composed of *on-Body*, *Body-to-Body* (or *inter-Body*) and *off-Body* communication networks. In inter-Body communications, a device coordinator with specific features could be responsible to communicate with the adjacent *Wireless Body Area Networks* (WBANs). A coordinator is generally considered as a resource-rich device, which can be a multi-standard node to interface with other technologies such as static *Wireless Sensor Networks* (WSNs), WiFi Access, Points or Broadband Cellular Networks (*e.g.*, 4G, LTE, *etc.*) [2]. The advent of such technology can thus be served as an ad-on to existing PSN infrastructures for improving the quality-of-service (QoS) and reliability.

In *Public Protection and Disaster Relief* (PPDR) scenarios, the network connectivity and data is a challenging problem due to the dynamic mobility and harsh environment [3]. It is envisioned that, in case of unavailable or out-of-range network infrastructures, the WBANs coordinators along with WBANs sensors can exploit cooperative and multi-hop body-to-body communications to extend the end-to-end network connectivity. In this regards, for on-body networks, the *Opportunistic and Mobility Aware Routing* (OMAR) scheme is proposed in one of our earlier works[4]. The emphasis of this study is on exploration of suitable routing protocols for *Body-to-Body Networks* (BBN) under dynamic mobility scenario. To the best knowledge of the authors, there is currently no comprehensive study using realistic disaster mobility model and scenario that could be considered as a benchmark. In the recent proposed works [5] also, routing protocols are evaluated without consideration to the communication technologies, therefore, we as a first step, investigating various classes of routing protocols to analyze their performance in BBN context with different communication technologies.

In this paper, based on the operational and technical requirements of the emergency and rescue operations [6], we investigate the behavior of multi-hop routing protocols with different wireless technologies for a fire emergency case inside a shopping mall. At first, a realistic mobility pattern and trace files are generated using a tool called BonnMotion [7]. These patterns and movements are geographically mapped and simulated over two incident areas inside the mall. The mobility trace file (which contains the nodes movements over space and time) is fed inside a packet-oriented network simulator called WSNet. This mobility model served as a realistic environment for the evaluation of various class of routing protocols. We considered *proactive*, *reactive*, *gradient-based* and *geographical-based* routing categories. The state-of-the-art individual protocol is selected among each category for a comparative study. For proactive, reactive, gradient-based and geographical-based routing protocols, *Optimized Link State Routing protocol* version 2 (OLSRv2) [8], *Ad-hoc On-Demand Distance Vector* (AODVv2) [9], *Directed Diffusion* (DD) [10] and *Greedy Perimeter Stateless Routing* (GPSR) [11] protocols are selected respectively. In addition, multi-standard technologies including IEEE 802.11 (WiFi), Zigbee / IEEE 802.15.4, and IEEE 802.15.6 (WBAN) are investigated for the Body-to-Body Network coordinator. Finally extensive network simulations are conducted and the results of average successful *Packet Reception Rate* (PRR), *Packet Delay* and *Energy Consumption* are presented.

The remainder of this paper is organized as follows. In Section II, we present the related works of several categories of routing protocols; while in Section III, we describe the target application and mobility scenario along with the target routing protocols. In Section IV, we introduce the simulation parameters and we present the obtained simulation results. Finally, in Section V, we conclude the paper and discuss future research directions.

II. RELATED WORKS

In addition to the technologies inter-operability, coexistence and energy consumption issues investigated in *Public Safety Networks (PSN)* [1], Routing is an important challenge to raise in critical and emergency operations. Since there is no dedicated routing protocols for PSN, researchers are tending to evaluate the performance of diverse routing protocols such as *Mobile Ad hoc Networks* (MANET), *Geographical Location based* or *Wireless Sensor Networks* (WSN) routing protocols. The following sections survey existing classes of routing protocols in PSN.

*A. Mobile Ad hoc Networks*

Due to its flexibility to topology changes and its multi-hop routing aspect, Mobile Ad hoc networks are interesting to be investigated in *PSN*. Mobile Ad hoc routing protocols are divided into four main classes: Proactive, Reactive, Hybrid and Hierarchical. These routing protocols refer commonly to the link-state and distance vector algorithms, hence their classification is based on network discovery and the routing information update mechanism [12].

Proactive routing protocols create and maintain continuously their routing tables, called also table-driven. In this routing class, nodes keep exchanging information to learn the network topology. Proactive protocols use one or more tables to store the topology information and routes. By these settled routes, optimization algorithms (such as *Dijkstra* in [13]) are applied to select best routes to use based on a selected metric. Proactive protocols differ on the technique used for neighbors sensing and topology update. The difference concerns also the messages used to: discover, maintain and share topology information and routes. Proactive routing protocols are appropriate with small networks. Due to the continuous broadcasted updates, routing overhead cause a considerable bandwidth consumption. The most widely known proactive routing protocols are*: Optimized Link State Routing protocol* (OLSR) [14], *Wireless Routing Protocol* (WRP), *Destination-Sequenced Distance-Vector* (DSDV) and recently the OLSR version 2 [8]. In *PSN*, such routes discovery techniques and computation cause network overloading that impact negatively the bandwidth utilization and increase the power consumptions. Drawbacks of proactive routing protocols in *PS* are not only limited into energy inefficiency and bandwidth overload, the routing convergence delay caused by the intensive routes discovery broadcasts at the network startup is also considerable.

Unlike proactive protocols, reactive routing protocols look up for routes only when it is needed. The route discovery procedure is invoked when data packets are ready for sending. The route discovery mechanism in reactive routing protocols is the same: a source node starts by flooding a request message to all reachable nodes looking up for the destination, then, each node relay this request message until it reaches the destination. If the destination is reached, a reply message is sent back to the source node through the reverse route followed by the request. Most known on-demand routing protocols: *Ad hoc On demand Distance Vector* (AODV) [15], *Dynamic Source Routing protocol* (DSR), *Temporally Ordered Routing Algorithm* (TORA) and more recently AODV version 2 (DYMO) [9]. Reactive routing protocols in Mobile Ad hoc networks seem to solve the bandwidth utilization and energy efficiency issues based on on-demand routing request, but the delay caused by the route discovery before data transfer does not meet the *PSN* requirements.

In addition to Ad hoc routing protocol classes mentioned above, hybrid routing protocols class merge both techniques: reactive and proactive routing protocols. These routing protocols offer generally a combination between proactive routing for nearby nodes and reactive routing protocols for far-away nodes. In *PSN*, this type of protocols could be appropriate for large catastrophe zone with many incident areas, where intra-incident area is proactive and inter-incident area is reactive routing. Another routing protocol class called hierarchical or clustering routing protocols propose to divide the entire network into groups of sub-networks in order to facilitate network management functionalities, especially the routing process. One of the most known protocol from this class called *Cluster Based Routing Protocol* (CBRP) is recently investigated in [16].

### B. Data-Centric Routing Protocols

A gradient routing protocol is interesting to investigate in specific cases of *PSN* where all the data flow converges towards only one node (*e.g.* command center). The existing gradient routing approaches are designed for Wireless Sensor Networks as data-centric routing. Quite different from traditional address-centric routing considered as a flat routing, where all nodes have the same interest and importance in the network. In data-centric routing, a sink node collects all data from the other nodes in two main steps. The sink node starts by broadcasting a request to all neighbors until it reaches the concerned node. The response follows back the request path. A node forwarding data may aggregate its own data with the traveling data towards the sink node. Various gradient routing technique exist, most are variants of *Directed Diffusion* [10].

### C. Geographical Location Based Routing Protocols

The drawbacks of Ad hoc networks in terms of continuous routes maintenance, storing of all network topology information into the nodes and network overload by unnecessary routes discovery (in case of proactive techniques), make further approaches come up to exceed these issues. Geographical based routing protocols are one of the proposed approaches. For more than ten years, geographical location based routing protocols avoided the technique of storing and sharing the network topology information. Routing decisions in geographic routing protocols are made hop-by-hop, no end-to-end routes made as in Ad hoc, for that, nodes in geographic routing protocol network store only physically reachable nodes information [17].

Hence, no routes maintenance needed, because packets could follow different paths each time. In *PSN*, geographic location is an important parameter required regardless the routing approach used. Outdoor geographic locations could be simply obtained based on GPS technology. Indoor localization could be based on anchor nodes or simply tags sending their location. However, basing the routing decisions on geographic location must rely on a high precision available location technology, otherwise, erroneous locations lead on inefficient routing. Most known routing protocols in this class, early in 2000, Greedy Perimeter Stateless Routing (GPSR) [11] followed by Geographic Source Routing (GSR) and Spatially Aware packet Routing (SAR) [18].

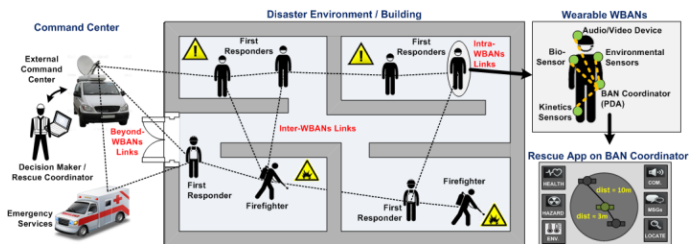

Fig. 1 On-Body and Body-to-Body Communications.

## III. MOTIVATIONS AND SCENARIO:

### A. Context and Motivations

This paper presents an ongoing research of the project Critical and Rescue Operations using Wearable Wireless sensor networks (CROW$^2$) [19]. The main objective of the project is to provide ubiquitous wireless communication and monitoring systems for emergency networks in disaster relief. It is anticipated that in emergency situation, the existing infrastructure network may be damaged (due to disaster itself), out-of-range or over saturated. In this realm, to extend the end-to-end network connectivity, WBANs coordinators can wirelessly interconnect the on-body sensors to external network infrastructures, by exploiting cooperative and multi-hop body-to-body or beyond-WBAN communications. Consequently, the on-field rescue members can be connected with external command center in midst of rescue operation.

Figure 1, illustrates the deployment and coordination of rescue teams in disaster environment which represent our rescue and emergency management application scenario. It is considered that each field team member is equipped with a set of wearable wireless sensor devices to monitor: i) its vital signs (*e.g.* heartbeats, blood pressure, stress level, *etc.*); ii) its motion and body posture (*e.g.* orientation, acceleration, heading, position, *etc.*); and iii) the surrounding environment (*e.g.* temperature, toxic gases, fire, light intensity, *etc.*). It is envisioned that these wearable WSNs will be heterogeneous in terms of supported data rates, sensors types, supported Quality of Services (QoS), communication ranges hardware capabilities, mobility and propagation conditions. Furthermore, these wearable WSNs can be deployed in harsh environments, thus requiring effective communication, coordination and monitoring capabilities.

### B. Mobility Modeling and Disaster Scenario

Evaluating and simulating a routing protocol in a specific context, strongly depends on the accurate mobility models. Mobility models metrics are classified as follows [20]:
- *Random based*: no dependencies or restriction.
- *Temporal dependencies*: current movements depend on the past ones.
- *Spatial dependencies*: movements depend on the movements of the surrounding units.
- *Geographical restrictions*: geographic restriction on the movements.
- *Hybrid structure*: Integration of two or more models.

An hybrid structured disaster area model designed by *Aschenbruck et al.* in [21] divides the catastrophe area into four different sub-areas. First, *incident site* contains one or more incident area(s) that represent(s) the exact incident location (*e.g.,* coordinates of aircraft crash, coordinates of a fire trigger, *etc.*). Second, *casualties treatment area* contains one or more patients waiting for treatment area and casualties clearing station. Then, the *transport zone* with ambulances and eventually rescue helicopter(s). The last sub-area is the *hospital zone*, which is often not represented, because size constraints, so arriving to the transport zone, casualties are considered cleared and safe. The last and important component in this model is the location of the *command center* responsible for conducting the rescue and emergency operations.

In this paper we investigate a disaster scenario (fire triggering as a case study) in the "Landmark" shopping mall in the State of Qatar as depicted in Figure 2. The mobility model used is generated by the *BonnMotion* tool. Let us first consider some logistic aspects for the mobility scenario. We consider that the incident is caused by a fire in two opposite sides in the mall (Figure 2). Then rescuers are called to react along with firefighters and medical teams. Firefighters are divided into 3 groups of vehicles with 26 firefighters in each group. Medical emergency teams that probably could reach the mall just after the incident, are consisting of 6 ambulances with 5 medical staff in each ambulance (30 personal in total).

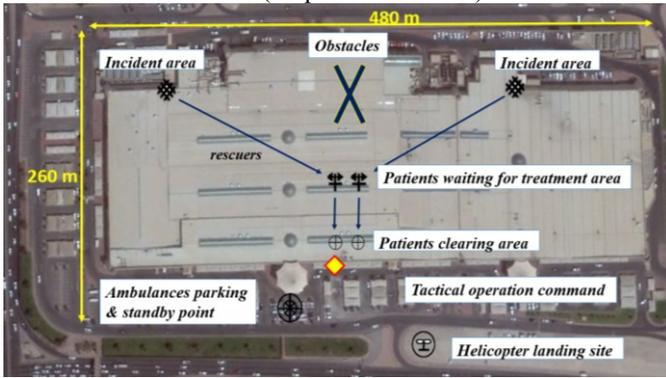

Fig. 2 Overview of the Disaster Scenario in the Landmark Shopping Mall

Further, police officers and civil defense personals are also considered to support the emergency teams (we have considered 18, to have a total of 100 rescuers). We consider all the rescuers as moving nodes and sending their gathered information to one main sink node placed at the main-gate of the mall (shown as ◆ in Figure 2). Data sent could be rescuers and/or victims health status, ambient rescuing conditions, special medical requests, *etc.*, based-on simple data, voice, images and/or video.

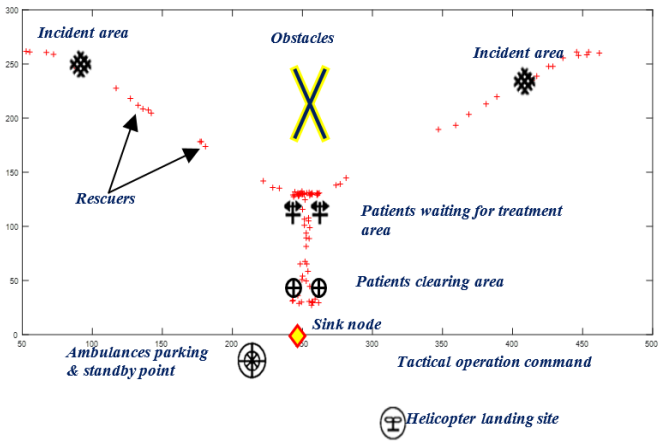

Fig. 3 Disaster Area Nodes Locations, Areas and Obstacles.

We provide, area perimeter coordinates, obstacles coordinates, number of nodes (*i.e.*, personal in our case) in each incident area, transported nodes in each group of nodes, *etc.*, as an input parameters to BonnMotion. As an output, we obtain a mobility trace file containing the movement of all the nodes during the observation time. The generated mobility trace file is used as an input for the comparative evaluation of the routing protocols (as illustrated in Figure 4) discussed in the following section.

### C. Routing approaches evaluated

#### 1) Proactive: OLSR version 2[8]

OLSRv2 is a proactive link state routing protocol that uses periodic local and global signaling for neighbor/link discovery and link state diffusion. Instead of OLSR (*i.e.*, first version), which uses four messages types, OLSRv2 require only two messages types: HELLO and TC (internal and external with only one parser for all messages types). It uses also addresses compression with IPv6 support. Three main steps are followed in this routing approach: neighborhood discovery, Multi-Point Relay (MPR) selection, routing table calculation and maintenance.

#### 2) Reactive: AODV version 2 [9]

In AODV, a node does not perform route discovery or maintenance until it is needed. A route discovery in AODV is initiated by the source node (S) through the broadcast of a specific route request (RREQ) to all its first hop neighbors. This route request is transferred by broadcasting until it reaches the destination. The RREQ records only the last_hop and the destination address.

AODV uses also the source and destination sequence number to detect and identify routes freshness. The RReq_ID is used to avoid processing duplicated requests (the pair Source_ID and RReq_ID is unique). A Time-To-Live (TTL) is also used by AODV to prevent an indefinite routing of a request. The Route Reply (RREP) in AODV follows the same route saved by the nodes while transferring the RREQ. This information will be deleted after a timer is expired if a RREP is not received. Besides, AODv2 records the entire traveled path into the RREQ packet header. It removes also the unnecessary RREP and Hello messages. AODVv2 is energy aware and selects routes based on the low energy index.

#### 3) Gradient based: Directed Diffusion[10]

Directed Diffusion (DD) is a data-centric routing protocol designed for WSN. With respect to WSN main requirements, DD is energy efficient, scalable and robust [22]. The routing mechanism in DD follows three steps. At startup, the sink node requests to gather data from one or more nodes. So, the sink node broadcasts the requests called interests towards the concerned node(s). Then, the routes (or gradients) are set up by selecting non-redundant route towards the sink node. This process starts with a low data rate specified by the sink node; afterwards, this data rate is reinforced by the sink node itself through one selected node. The *reinforcement* then is propagated throughout all the nodes.

#### 4) Geographic based: GPSR [11]

GPSR uses the nodes location and the wireless connectivity. It uses two forwarding techniques: greedy forwarding and perimeter forwarding. In greedy forwarding, packets from source node to destination are forwarded throughout the geographically closest next hop towards the destination. When a greedy forwarding is impossible, the protocol routes the packets in the surrounding perimeter of the destination. GPSR returns to the greedy forwarding early when a local maxima (local parameter) is reached. GPSR maintains only its location and locations of neighbors.

## IV. PERFORMANCE EVALUATION

### A. Simulations setup

In this section, proactive (*i.e.*, OLSRv2), reactive (*i.e.*, AODV), geographic based (*i.e.*, GPSR) and gradient based (*i.e.*, DD) routing protocols are evaluated using a realistic disaster mobility model as explained in Section III.B. In addition, we consider various communication technologies including WiFi (*i.e.*, IEEE 802.11 standard), WBAN (*i.e.*, IEEE 802.15.6 standard) and WSN (*i.e.*, IEEE 802.15.4 standard). These wireless technologies (*i.e.*, MAC and PHY layers), are selected especially to analyze and evaluate inter-body communication (*i.e.*, realized through WBAN coordinator). Subsequently, each of these technologies is implemented using above selected routing protocols for comprehensive evaluation. We are using an event-driven, packet-oriented network simulator called *WSNet (version 3.0)*, for systems level simulations. The simulations are executed based on a realistic mobility model for 100s. We considered 10 iterations for every simulation and the 95% confidence intervals are provided. All the parameters at each layer are configured though an XML configuration file, where we vary the routing protocols for each technology.

TABLE 1. LIST OF SIMULATIONS PARAMETERS AND CORRESPONDING VALUES

| Standards | MAC Layer | PHY Layer | Battery Parameters (mA) | | |
|---|---|---|---|---|---|
| | | | TX | RX | IDLE |
| WiFi IEEE 802.11 | CSMA/CA DCF with ACK | Modulation BPSK, Sensitivity = -92dBM, TX Power = 0dBm, 2.4GHz | 160 | 53 | 0.69 |
| WSN IEEE 802.15.4 | CSMA/CA without ACK | Modulation O-QPSK, Sensitivity = -85dBM, TX Power = 0dBm, 2.4GHz | 17.4 | 19.7 | 0.9 |
| WBAN IEEE 802.15.6 | CSMA/CA with ACK | Modulation DQPSK, Sensitivity = -85dBM, TX Power = 0dBm, 2.4GHz | 17.4 | 19.7 | 0.9 |

Concerning the performance metrics, we consider the *Packet Reception Rate* (PRR), *Communication Delay* and *Energy Consumption* as the main metrics. The complete simulation process as shown in Figure 5 is a set of operations iterated for 10 times to converge to the realistic behavior of the evaluated routing protocols. At first, the mobility generation tool BonnMotion with specific parameters of the studied disaster scenario generates Mobility trace file. The output file is then converted into a proper format before being parsed by the simulator WSNet. A routing protocol is then selected with a specific communication technology and an initial payload. These parameters are entered through the "XML" configuration file.

#### 1) Application and Routing Layers

At the application layer, we consider 99 moving nodes (*i.e.* WBANs coordinators) inside the shopping mall sending data packets to one sink node (*i.e.* command center), here node 0, which is placed at the main gate of the mall. Distance between nodes, movements, directions and speed are calculated according to the mobility model. A *Constant Bitrate Rate* (CBR) application is used to generate the traffic (with one packet/s), with available data payload ranging from 2 bytes to 256 bytes. At the network layer, a routing protocol detailed in Section III-C is selected as illustrated in the simulation process (*cf.* Figure 5). The routing layer receives the packets from the application layer, depending on the routing protocol; all the configuration parameters are equally affected. Each routing approach will be evaluated with individual technology detailed in Table 1.

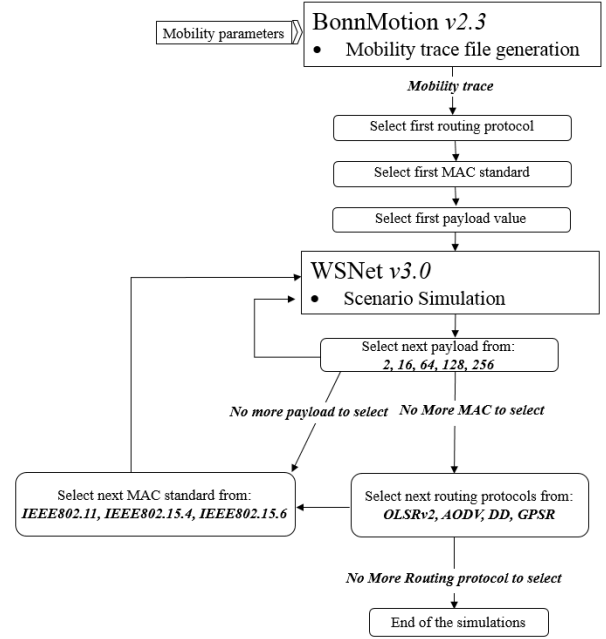

Fig. 4. Simulation Methodology.

#### 2) MAC and Physical Layers

At the MAC layer, we are employing unified distributed CSMA/CA protocol for the three wireless technologies. It includes *DCF IEEE 802.11* (for *Wifi*) which employs a CSMA/CA with binary exponential back-off algorithm. It uses CTS/RTS control signals for better reliability. *IEEE 802.15.4-based CSMA/CA (*for *WSN)* is implemented with maximum back-off exponent set as 3; maximum back-off is 5 without any re-transmission.

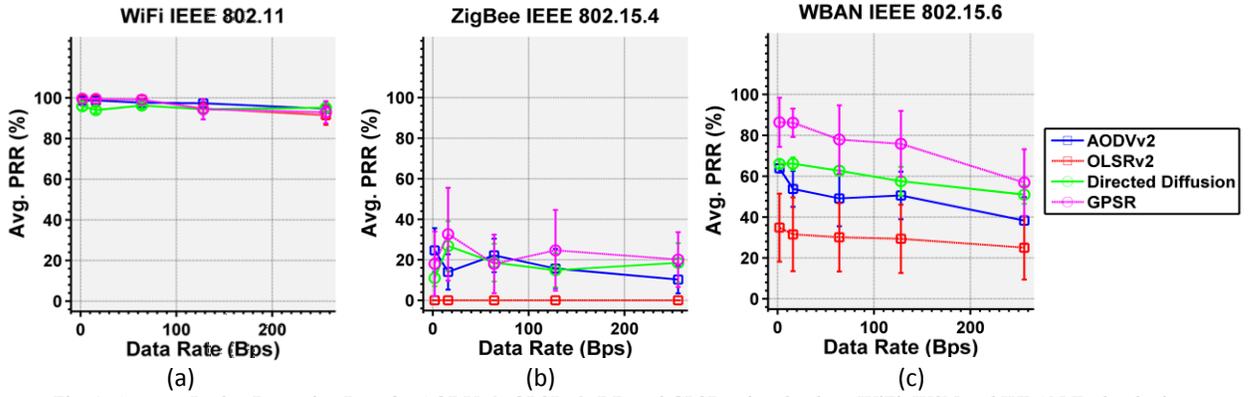

Fig. 1. Average Packet Reception Rate for AODVv2, OLSRv2, DD and GPSR using the three WiFi, WSN and WBAN Technologies.

Finally, *IEEE 802.15.6 (for WBAN) CSMA/CA* MAC protocol with immediate acknowledgment policy is implemented. We have exploited the higher emergency level feature of this standard (*i.e.* 2) for the transmitted packets. The maximum back-off is set as 5 and re-transmission limit is 3. Along with the selected MAC layers, corresponding modulation schemes, physical configuration parameters including transmit power levels and corresponding current consumptions (of the widely used radio transceivers *i.e.*, cc2420 for WSN/WBAN, and cc3100 for WiFi) are detailed for various states in Table 1.

*B. Simulation Results*

In this section, the performance of OLSRv2, AODVv2, GPSR and DD are investigated with the WiFi, WSN and WBAN technologies (*i.e.* IEEE 802.11, IEEE 802.15.4 and IEEE 802.15.6 standards). Parameters settings were configured for the context of *PSN*.

*1) Packet Reception Rate (PRR)*

The results of average PRR for the four selected protocols and three technologies are shown in Figure 4. Generally, the evaluated routing protocols perform much better using WiFi in comparison to the others technologies, overall it achieves more than 92% of PRR. In particular, with low payloads (*i.e.* 2 and 16 bytes), only DD PRR is below 96% whereas, OLSRv2, AODVv2 and GPSR are all able to achieve above 99% PRR. Starting from 64 bytes and higher payload, DD performance also starts improving to exceed others protocol performance as can be seen in Figure 4-a. GPSR and OLSRv2 showed similar performance. AODVv2 has best performance with *WiFi* at 128 bytes payload, and performs similar as DD with 256 bytes payload. Finally, GPSR shows slightly lower performance with more than 128 bytes payload. This is due to the perimeter forwarding technique which may occur several times due to the obstacles located in the mobility model.

For the case of *WSN IEEE 802.15.4 (ZigBee)*, in general, all the protocols are under-performed as shown in Figure 4-b. As the best case, 33% of average PRR was achieved using GPSR. Even with the lowest payloads, the performance remained very low. OLSRv2 was not able to deliver any packet at all with the various payload values. *WSN* is a short-range communication technology, with high mobility nodes such as defined in the mobility model and according to the nodes density in the shopping mall; evaluated routing protocols are unable to perform well with *WSN*. Additionally, by using CSMA/CA MAC without any acknowledgment policy the performance further degrades.

*WBAN (i.e., IEEE 802.15.6)* is mainly an intra-BAN communication technology, but recently, research trends are tend to evaluate this standard in inter-BAN context[23]. For this reason, we are investigating *IEEE 802.15.6* standard to achieve the potential limits studied in [2]. In *WBAN* generally, most of the protocols perform much better in comparison to *WSN* technology. In particular, GPSR outperforms the other routing approaches again, it achieves up to 88% PRR under low payloads (2 and 16 bytes). However, by increasing the payload, GPSR starts to gradually degrade in performance same as the case in other technologies however,GPSR remains the best protocol. OLSRv2 has the worst performance, whereas both DD and AODVv2 also reach below 50% average PRR with 256 bytes. As we analyzed the performance given by combining one of the evaluated routing protocols with *WBAN IEEE 802.15*.6, GPSR meets the disaster scenario requirements with low payload, the rest of protocols are inconclusive. It is necessary to notice that while considering WBAN technology, the low data rate is a limitation in terms of image and video transfer.

Finally, the evaluated routing protocols used with *WiFi* are convincingly better than the two counterparts in terms of average PRR. Only, GPSR performed well with *WBAN*.

*2) Latency:*

We considered latency as the average packet delay between the source node and the final destination over a multi-hop BBN. Generally, the results of the delay are inter-related with PRR, if PRR is higher then, delay will be lower. Focusing onto *WiFi* technology, with low payloads (*i.e.* 2 and 16 bytes), all routing protocols delay is below 80ms which satisfy our application context. Figure 6-a shows an exponential increase in delay for AODVv2 and GPSR starting from 16 bytes of payload, while OLSRv2 delay remains slightly lower than 80ms until 16 bytes. DD is the most efficient and has almost negligible delay among all protocols and therefore is considered as the most effective protocol in terms of delay using *WiFi* technology.

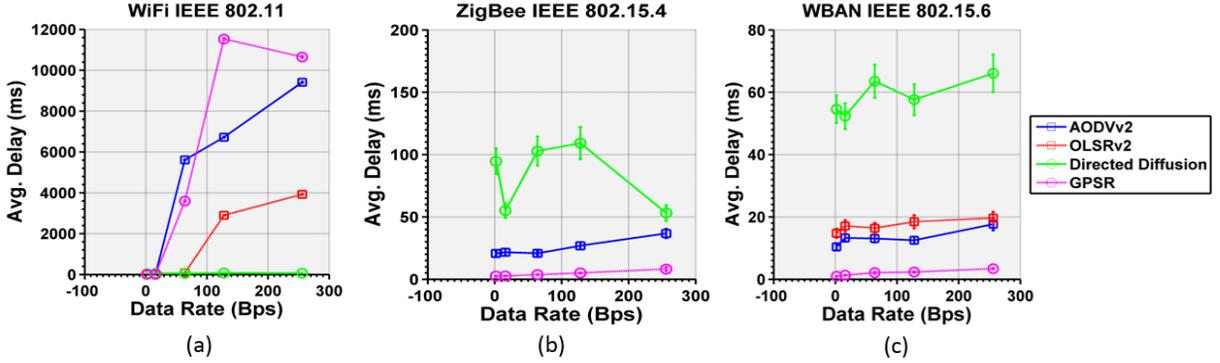
Fig. 2. Average Communication Delay for AODVv2, OLSRv2, DD and GPSR using the three WiFi, WSN and WBAN Technologies.

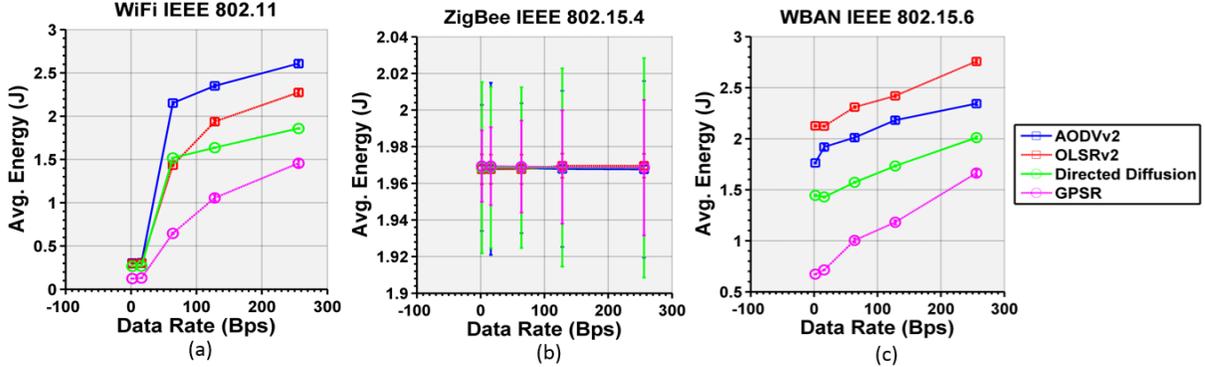
Fig. 3. Average Energy Consumption for AODVv2, OLSRv2, DD and GPSR using the three WiFi, WSN and WBAN Technologies.

In comparison to the other evaluated protocols, DD has very low calculation for data routing. While, AODVv2 with an on-demand routes lookup technique and GPSR which also bases its routing table on geographic locations calculations which require more time to route calculation before data transfer which adds an additional significant delay. Figure 6-b, shows the average delay for WSN. It is clear that comparatively, it is extremely (*i.e.*, 100 times) better than *WiFi*. AODVv2, DD and GPSR all perform very well even at higher payloads sizes. Exceptionally, OLSRv2 performance is the worst one since it was unable to transmit any packet and had zero average PRR, therefore the delay is irrelevant for it. Although, DD shows slightly more delay comparing to the other two protocols, but overall it is below 100ms as well.

With *WBAN* technology, in addition to the respectable PRR recorded by GPSR with low payload, all the routing approaches perform much better in term of delay with WBAN technology than with the rest of the communication technologies. Results in Figure 6-c, show that delays are very low in comparison to the delays recorded with *WiFi* and *WSN* technologies.

Finally, concerning the delay, *WBAN* and *WSN* outperformed *WiFi* in most of the protocols. Only in case of GPSR with *WBAN*, the results are comparable and it is the most effective protocol for optimized delay performance.

*3) Energy Consumption:*

The energy consumption for each transmitted packet is calculated as follows,
$$E_{packet} = T_{packet} \times 3_{Volts} \times I_{mA}$$
where, $T_{packet}$ is the duration in ms which is based on the effective packet length (including all the *PHY and MAC headers*[2]). The current consumption values for two different considered radio transceivers are mentioned in Table 1. For *WiFi*, approximately linear increase in energy consumption is observed with an increase in payload size for all the protocols as shown in Figure 7–a. There is hardly any difference between the protocols for 2 and 16 bytes of payload. However, for higher payloads it is notable that AODVv2 consumes the highest energy, whereas GPSR is the most energy efficient protocol.

In general, all routing protocols have performed much better with *WSN* for energy efficiency. By increasing the payloads, the energy consumption decreases for AODVv2, DD and GPSR. It seems that OLSRv2 performed very well in terms of energy consumption, but since it does not able to transmit any data packet, this reflection is not for a successful transmission and hence irrelevant In addition, the behavior of OLSRv2 is a reflection of continuous routes maintenance operations, which increases energy consumption even without packet receptions. Despite of that OLSRv2 improves OLSR features, though, it still not scalable and destined for small networks with small data traffic. Using *WBAN*, routing protocols, except AODVv2 and GPSR have similar pattern, *i.e.*, energy consumption slightly decrease with the increase in payload. OLSRv2 performs better than the other routing protocols. A slight difference of energy consumption is noticed between the evaluated routing protocols. Finally, in terms of energy efficiency, *WBAN* communication technology is most suitable with the evaluated routing protocols.

To summarize, definitely *WiFi* is the most relevant for the reliable communication but at an expense of significant increase in delay. In this aspect *WBAN* technology is most effective and by using GPSR routing protocol it can be a considerable option for *BBN*. GPSR with both *WiFi* and *WBAN* is able to achieve high packet reception and consumes relatively much lower energy with low delay. Referring to our network topology (*i.e.*, converge cast) and mobility scenario, GPSR is one of the most favorable protocol as reflected in Table 2. Finally, for small-scale rescue and critical operations using BBN, both WiFi and WBAN can be considered based on the given constraint, either PRR or delay.

TABLE 2. COMPARATIVE TABLE FOR ROUTING PROTOCOLS BEHAVIOR WITH DIFFERENT WIRELESS TECHNOLOGIES

| Standards | Routing | PRR | Delay | Energy |
|---|---|---|---|---|
| WiFi IEEE 802.11 | AODVv2 | High | High | High |
| | OLSRv2 | High | High | Medium |
| | DD | Medium | Low | Medium |
| | **GPSR** | **High** | High | **Low** |
| WSN IEEE 802.15.4 | AODVv2 | Low | Low | **Low** |
| | OLSRv2 | Worst | Worst | Worst |
| | DD | Low | High | Low |
| | GPSR | Low | Low | Low |
| WBAN IEEE 802.15.6 | AODVv2 | Low | Low | Low |
| | OLSRv2 | Low | Low | **Low** |
| | DD | Medium | Medium | Low |
| | **GPSR** | **High** | **Low** | **Low** |

## V. CONCLUSION

Although in recent years, broadband networks such as LTE has emerged as viable option for PSN, but there is always a potential risk that such infrastructure-based network could be damaged or oversaturated during the disaster and emergency situations. In this context, Wearable Wireless Sensor Networks could not only be served as an add-on to existing PSN but can also help to monitor the critical health and environment status during the rescue and evacuation process. In this paper, a particular emphasis is given to the emerging body-to-body communication whilst evaluating best wireless technologies and routing strategies under realistic mobility scenario for public safety and disaster relief operations. Three technologies (*i.e.*, WiFi IEEE 802.11, WSN IEEE 802.15.4 and WBAN IEEE 802.15.6) and four different class of routing protocols are considered including mobile Ad hoc (*i.e.*, OLSRv2 and AODVv2), data centric (directed diffusion) and geographical location-based (GPSR). It is concluded that WiFi is the best technology for both packet reception ratio and energy efficiency performance metric. Whereas, as far as the packets delay is concerned, WBAN is the most effective technology. Among the protocols, by assuming that we have location information, then GPSR performed the best in comparison to all other protocols using WiFi IEEE 802.11. The only exception is with delay results, where DD outperformed all other protocols. If location information is not available, then DD especially with WBAN IEEE 802.15.6 can be considered as a favorable choice. However, it is important to note that WBAN has maximum payload limit of 256 bytes which limits it to the transmission of real-time audio or video.

## ACKNOWLEDGMENT

This publication was made possible by NPRP grant # [6-1508-2-616] from the Qatar National Research Fund (a member of Qatar Foundation). The statements made herein are solely the responsibility of the authors.

## REFERENCES

[1] M. Alam and E. B. Hamida, "Surveying Wearable Human Assistive Technology for Life and Safety Critical Applications: Standards, Challenges and Opportunities," *Sensors*, vol. 14, no. 5, pp. 9153–9209, May-2014.
[2] M. M. Alam, D. B. Arbia, and E. B. Hamida, "Device-to-Device Communication in Wearable Wireless Networks," *10th CROWNCOM Conf.*, Apr-2015.
[3] E.B Hamida and G. Chelius, "Analytical Evaluation of Virtual Infrastructures for Data Dissemination in Wireless Sensor Networks with Mobile Sink", In Proceedings of the ACM MOBICOM First Workshop on Sensor Actor Networks, Montreal, Canada, Sept-2007
[4] E.B Hamida, M. M. Alam, M. Maman, and B. Denis, "Short-Term Link Quality Estimation for Opportunistic and Mobility Aware Routing in Wearable Body Sensors Networks," *WIMOB 2014 2014 IEEE 10th Int. Conf. Wirel. Mob. Comput. Netw. Commun. WiMob*, p. pp: 519–526, Oct-2014.
[5] "D. G. Reina, M. Askalani, S. L. Toral, F. Barrero, E. Asimakopoulou, and N. Bessis, 'A Survey on Multihop Ad Hoc Networks for Disaster Response Scenarios,' International Journal of Distributed Sensor Networks, Article ID 647037, in press.," May-2015.
[6] National Public Safety Telecommunications Council, "A NPSTC Public Safety Communications Report - Defining Public Safety Grade Systems and Facilities," National Public Safety Telecommunications Council, May-2014.
[7] N. Aschenbruck, "BonnMotion :A Mobility Scenario Generation and Analysis Tool." University of Osnabruuck, Jul-2013.
[8] T. Clausen, C. Dearlove, P. Jacquet, and U. Herberg, "RFC7181: The Optimized Link State Routing Protocol Version 2" Apr-2014.
[9] C. Perkins, S. Ratliff, and J. Dowdell, "Dynamic MANET On-demand (AODVv2) Routing draft-ietf-manet-dymo-26." Feb-2013.
[10] C. Intanagonwiwat, R. Govindan, and D. Estrin, "Directed diffusion: a scalable and robust communication paradigm for sensor networks,"pp. 56–67,2000.
[11] B. Karp and H. T. Kung, "GPSR: Greedy Perimeter Stateless Routing for Wireless Networks," *Annu. ACMIEEE Int. Conf. Mob. Comput. Netw. MobiCom 2000*, no. 6, 2000.
[12] M. Abolhasan, T. Wysocki, and E. Dutkiewicz, "A review of routing protocols for mobile ad hoc networks," *Ad Hoc Netw.*, vol. 2, no. 1, pp. 1–22, Jan-2004.
[13] J. Yi, E. Cizeron, S. Hamma, B. Parrein, and P. Lesage, "Implementation of multipath and multiple description coding in olsr," *ArXiv Prepr. ArXiv09024781*, 2009.
[14] T. Clausen and P. Jacquet, "OLSR, IETF, RFC3626." Oct-2003.
[15] C. E. Perkins and E. M. Royer, "Ad-hoc on-demand distance vector routing," in *Mobile Computing Systems and Applications, 1999. Proceedings. WMCSA'99. Second IEEE Workshop on*, pp. 90–100, 1999.
[16] L. E. Quispe and L. M. Galan, "Behavior of Ad Hoc routing protocols, analyzed for emergency and rescue scenarios, on a real urban area," *Expert Syst. Appl.*, vol. 41, no. 5, pp. 2565–2573, Apr-2014.
[17] F. Cadger, K. Curran, J. Santos, and S. Moffett, "A Survey of Geographical Routing in Wireless Ad-Hoc Networks," *IEEE Commun. Surv. Tutor.*, vol. 15, no. 2, pp. 621–653, 2013.
[18] S.-H. Cha, "A Survey of Greedy Routing Protocols for Vehicular Ad Hoc Networks," *Smart Comput. Rev.*, vol. 2, no. 2, Apr-2012.
[19] E. B. Hamida, M. M. Alam, M. Maman, B. Denis, R. D'Errico "Wearable Body-to-Body Networks for Critical and Rescue Operations - The CROW[2] Project, IEEE 25th Annual International Symposium on Personal, Indoor, and Mobile Radio Communications (IEEE PIMRC 2014), Workshop on The Convergence of Wireless Technologies for Personalized Healthcare, Washington DC", Sept-2014.
[20] S. Kumar, S. C. Sharma, and B. Suman, "Mobility Metrics Based Classification & Analysis of Mobility Model for Tactical Network," *Int. J. -Gener. Netw.*, vol. 2, no. 3, pp. 39–51, Sep-2010.
[21] N. Aschenbruck, E. Gerhards-Padilla, M. Gerharz, M. Frank, and P. Martini, "Modelling mobility in disaster area scenarios," in *Proceedings of the 10th ACM Symposium on Modeling, analysis, and simulation of wireless and mobile systems*,pp. 4–12, 2007.
[22] S. K. Singh, M. . Singh, and D. K. Singh, "Routing Protocols in Wireless Sensor Networks - A Survey," *Int. J. Comput. Sci. Eng. Surv.*, vol. 1, no. 2, pp. 63–83, Nov-2010.
[23] M. M. Alam and E. B. Hamida, "Interference Mitigation and Coexistence Strategies in IEEE 802.15.6 based Wearable Body-to-Body Networks," in *10th CROWNCOM Conference*, Apr-2015.